\documentclass[preprint,amsmath,amssymb,aps,amsart,floatfix]{article}

\usepackage[title]{appendix}

\usepackage{graphics}
\usepackage{amsmath,amssymb,amsfonts}
\usepackage{mathtools}
\usepackage{hyperref}

\usepackage{url}
\usepackage{mathtools}
\usepackage{authblk}
\usepackage[bottom]{footmisc}

\begin{document}
\title{A unified view of the first-excited $2^+$ and $3^-$ states of Cd, Sn and Te isotopes}
\author[1,\footnote{Email: bhoomika.physics@gmail.com (Bhoomika Maheshwari)}]{Bhoomika Maheshwari }

\affil[1]{ University of Malaya, Kuala Lumpur, Malaysia.  }

\maketitle

\begin{abstract}
Symmetries are known to play an important role in the low lying level structure of Sn isotopes, mostly in terms of the seniority and generalized seniority schemes. In this paper, we revisit the multi-j generalized seniority approach for the first excited $2^+$ and $3^-$ states in the Cd ($Z=48$), Sn ($Z=50$) and Te ($Z=52$) isotopes, where the Cd and Te isotopes represent two-proton hole and two-proton particle nuclei, thus involving both kind of particles (protons and neutrons) in contrast to Sn isotopes. Interestingly, the approach based on neutron valence space alone is able to explain the B(E2) and B(E3) trends respectively for the $2^+$ and $3^-$ states in all the three Cd, Sn and Te isotopes. The new results on the inverted parabolic behavior of B(E3) values in Cd and Te isotopes are understood in a manner identical to that of Sn isotopes by using the generalized seniority scheme. No shell quenching is supported by these calculations; hence, the neutron magic numbers, $N=50$ and $N=82$, remain robust in these isotopic chains. It is quite surprising that the generalized seniority continues to be reasonably successful away from the semi-magic region, thus providing a unifying view of the $2^+$ and $3^-$ states.
\end{abstract}

\section{Introduction}

Symmetries provide deep theoretical insights of the structure of matter and fundamental forces, and offer a unified view of many phenomena. The connection between symmetries and conservation laws plays a key role in this respect. For example, conservation laws for energy and momentum can be understood due to the  symmetries with respect to translations in time and space. The first studied symmetries in nuclear physics include charge independence and charge symmetry, which led to the concept of isospin, first introduced by the Nobel Prize winning nuclear theorist, Eugene Wigner~\cite{wigner}. Many nuclear structure phenomena reflect the symmetries of the basic building blocks and their interactions, and serve as the laboratories for studying the symmetries of the fundamental processes themselves ~\cite{bohr1975}. In this paper, we invoke the symmetry of the shell model ~\cite{mayer1955} related to the pairing interaction in terms of the goodness of generalized seniority quantum number ~\cite{racah1943,shalit1963,casten1990,heyde1990,talmi1993,isacker2014,bhoomika,kota}.  

Our recent studies of the seniority scheme and the generalized seniority scheme~\cite{maheshwari2016,maheshwari20161,jain2017,jain20171,maheshwari2017,maheshwari2019,maheshwari20192} has offered a simplified view of the complex structure of various states in semi-magic nuclei and successfully explained a number of spectroscopic properties like the energy spectra, B(EL) trends, half-lives, g-factor trends, Q-moments etc. A new kind of seniority isomers in Sn isotopes has been found and related generalized seniority selection rules were quoted \cite{maheshwari2016}. The role of multi-j configuration is found to be essential. Such suggested multi-j configurations for a given state remain consistent in explaining various properties from reduced transition probabilities to moments. The explanation for the first $2^+$ states in Sn isotopes were made \cite{maheshwari20161} which led to the similar understanding of first $2^+$ states in Cd and Te isotopes \cite{maheshwari20192}. In this paper, we review the generalized seniority results of first $2^+$ states for not only semi-magic nuclei but also away from semi-magic nuclei, particularly for the Cd isotopes (2-proton holes) and the Te isotopes (2-proton particles) in comparison to the Sn isotopes (Z=50 closed shell). We also extend the study of first $3^-$ states in Sn isotopes \cite{maheshwari2017}, in the chains of Cd and Te isotopes, for the first time. Presence of both protons and neutrons in the valence space makes the study very interesting for the first excited $2^+$ and $3^-$ states in all the three Cd, Sn and Te isotopic chains.   

We have divided the paper into five sections. Section 2 presents some essential theoretical details and key features of the generalized seniority scheme. Section 3 provides an empirical test to the generalized seniority. Section 4 discusses the calculated results for B(EL; L=2 or 3) trends of the first excited $2^+$ and $3^-$ states in Cd, Sn and Te isotopes and compares it with the experimental data wherever available. Section 5 summarizes the work.  

\section{\label{sec:level2}Theoretical formalism and key features}

\subsection{Seniority and Generalized Seniority}

Seniority scheme is usually credited to Racah~\cite{racah1943} but Flowers~\cite{flowers1952} also introduced it almost simultaneously though independently. Seniority ($v$) may be defined as the number of unpaired nucleons for a given state ~\cite{shalit1963,casten1990,heyde1990}. The quasi-spin scheme of Kerman~\cite{kerman1961} and Helmers~\cite{helmers1961}, for identical nucleons in single-j shell satisfies the SU(2) algebra formed by the pair operators $S_j^+$, $S_j^-$ and $S_j^0$ and is well suited to describe the seniority scheme, where 
$S_j^+= \frac{1}{2} {(-1)}^{(j-m)} a_{jm}^+ a_{j,-m}^+ $ , 
$S_j^-= \frac{1}{2} {(-1)}^{(j-m)} a_{j,-m} a_{jm} $ and $S_j^0= \frac{1}{2} (n_j-\Omega) $. Here $n_j$ and $\Omega$ are the number operator and pair degeneracy of the single-j shell. The details of this algebra and corresponding selection rules can be found in the book of Talmi~\cite{talmi1993}. The beauty of seniority lies in the reduction of $n$ nucleons (identical) problem to the $v$ nucleons problem by transforming the reduced matrix elements in $j^n$ configuration to the reduced matrix elements in $j^v$ configuration.

When $n$ identical nucleons occupy a multi-j space then the corresponding reduced matrix elements can be calculated in the generalized seniority scheme, which was first introduced by Arima and Ichimura~\cite{arima1966} for many degenerate orbitals. The corresponding generalized pair creation operator can be defined as $S^+ = \sum_{j} S^+_j$, where the summation over $j$ takes care of the multi-j situation~\cite{talmi1993}. Such generalized pair operators also satisfy the SU(2) algebra. Talmi further incorporated the non-degeneracy of the multi-j orbitals by using $S^+ = \sum_{j} \alpha_j S^+_j$, where $\alpha_j$ are the mixing coefficients~\cite{talmi1971,shlomo1972}. Our recent extension of this scheme for multi-j degenerate orbitals by defining $S^+ = \sum_j {(-1)}^{l_j} S^+_j$, as proposed by Arvieu and Moszokowski~\cite{arvieu1966}, led to a new set of generalized seniority selection rules ~\cite{maheshwari2016,maheshwari20161,jain2017,jain20171,maheshwari2017,maheshwari2019,maheshwari20192}. Here $l_j$ denotes the orbital angular momentum of the given-j orbital. An important non-trivial consequence was the discovery of a new category of seniority isomers, decaying via odd tensor transitions ~\cite{maheshwari2016}. The seniority in single-j changes to the generalized seniority $v$ in multi-j with an effective-j defined as ${\tilde{j}}= j \otimes j^\prime....$ having a pair degeneracy of $\Omega= \sum_j \frac{2j+1}{2} = \frac {(2 {\tilde{j}}+1)} {2}$. The shared occupancy in multi-j space is akin to the quasi-particle picture. However, the number of nucleons $n=\sum_j n_j$ and the generalized seniority $v=\sum_j v_j$ remain an integer. In this paper, we show the generalized seniority scheme with quasi-spin operators as $S^+ = \sum_j {(-1)}^{l_j} S^+_j$ ~\cite{arvieu1966}, where $S_j^+ = \sum_m {(-1)}^{j-m} a_{jm}^+ a_{j,-m}$ ~\cite{talmi1993}. A simple pairing Hamiltonian in multi-j space can hence be defined as $H= 2 S^+ S^- $, with the energy eigen values $[2s(s+1)-\frac{1}{2} (\Omega-n)(\Omega+2-n)]$ $= \frac{1}{2} [(n-v) (2 \Omega+2-n-v)]$. Here, $s=\sum_j s_j$ is the total quasi-spin of the state. 

Kota~\cite{kota2017} has recently shown that for each set of $\alpha_j$ values ($= {(-1)}^{l_j}$), there exists a corresponding symplectic algebra Sp(2$\Omega$) arising from U(2$\Omega$) $\supset$ Sp(2$\Omega$) with $\Omega= \sum_j \frac{2j+1}{2}$. This one-to-one correspondence between Sp(N) $\leftrightarrow$ SU(2) leads to special selection rules for electro-magnetic transition operators connecting $n-$nucleon states having good generalized seniority, which are in agreement with the selection rules obtained by us ~\cite{maheshwari2016}. 

\subsection{Key features}

The most prominent signatures of good seniority states show up in the behaviour of the excitation energy, electromagnetic transition rates like B(EL) and B(ML) values, Q-moments and magnetic moments, or g-factor values. We can summarize the features of good generalized seniority states as follows: 

\begin{itemize}

\item The excitation energies of good generalized seniority states are expected to have a valence particle number independent behaviour, similar to the good seniority states arising from single-j shell. It is rather easy to show this by extending the proof for the single-j seniority scheme by defining a multi-j effective configuration as $\tilde{j}$, as shown in \cite{maheshwari20192}. Consequently, the energy difference remains independent of the valence particle number for a given multi-j configuration. For example, the first excited $2^+$ states in Sn isotopes are observed at nearly constant energy throughout the isotopic chain from N=52 to 80.

\item The magnetic dipole moments i.e. g-factors for a given generalized seniority state show a constant trend with respect to particle number variation. In general, the magnetic transition probabilities support a particle number independent behaviour for both the even and odd multipole transitions. 

\item The electric transition probabilities exhibit a parabolic behaviour for both the odd and even multipole transitions. We recall these developments by the following expressions for electric multipole $L $(even or odd) operators as:\\
(a)	For generalized seniority conserving ($\Delta v=0$) transitions 
\begin{eqnarray}
\langle {\tilde{j}}^n v l J ||\sum_i r_i^L Y^{L}(\theta_i,\phi_i)|| {\tilde{j}}^n v l J \rangle = \Bigg[ \frac{\Omega-n}{\Omega-v} \Bigg] \langle {\tilde{j}}^v v l J ||\sum_i r_i^L Y^{L}(\theta_i,\phi_i)|| {\tilde{j}}^v v l J \rangle 
\end{eqnarray}
(b)	For generalized seniority changing ($\Delta v=2$) transitions
\begin{eqnarray}
\langle {\tilde{j}}^n v l J || \sum_i r_i^L Y^{L}(\theta_i,\phi_i)|| {\tilde{j}}^n v\pm 2 l J \rangle  = 
  \Bigg[ \sqrt{\frac{(n-v+2)(2\Omega+2-n-v)}{4(\Omega+1-v)}} \Bigg]\nonumber \\ \langle {\tilde{j}}^v v l J ||\sum_i r_i^L Y^{L}(\theta_i,\phi_i)|| {\tilde{j}}^v v\pm 2 l J \rangle 
\end{eqnarray} 

\item For $L=2$, Eq.(1) can directly be related to the electric quadrupole moments $ Q= \langle {\tilde{j}}^n J ||\hat{Q}|| {\tilde{j}}^n J \rangle= \langle {\tilde{j}}^n J ||\sum_i r_i^2 Y^{2}|| {\tilde{j}}^n J \rangle $ with the following conclusions: The Q-moment values depend on the pair degeneracy ($\Omega$), particle number ($n$) and the generalized seniority ($v$) as per the square bracket shown in Eq.(1). The $Q-$ moment values follow a linear relationship with $n$. The Q-values change from negative to positive on filling up the given multi-j shell with a zero value in the middle of the shell due to $\frac{\Omega-n}{\Omega-v}$ term. This is in direct contrast to the Q-moment generated by collective deformation which is expected to be the largest in the middle of the shell. 

\item The dependence of the $\sqrt{B(E2)}$ with particle number $n$ in Eq.(2) for the generalized seniority changing transitions is different than the case of $Q-$ moments. The $\sqrt{B(E2)}$ values for $\Delta v=2$ transitions exhibit a flat trend throughout the multi-j shell, decreasing to zero at both the shell boundaries. A nearly spherical structure is supported at both the ends for the given multi-j shell.

\item The first excited $2^+$ states with generalized seniority $v=2$ usually decays to the ground $0^+$ states, a fully pair-correlated state with generalized seniority $v=0$. Such $E2$ transitions are the generalized seniority changing transitions ($\Delta v=2$), where the corresponding B(E2) values can be obtained as follows: 
\begin{eqnarray}
B(E2)=\frac{1}{2J_i+1}| \langle {\tilde{j}}^n v l J_f || \sum_i r_i^2 Y^{2}(\theta_i,\phi_i) || {\tilde{j}}^n v\pm2 l' J_i \rangle |^2
\end{eqnarray}
The involved reduced matrix elements can similarly be obtained by using Eq.(2) between initial $J_i$ and final $J_f$ states with respective parities of $l$ and $l'$.     

\item Similarly, the first excited $3^-$ states with generalized seniority $v=2$ usually decays to the ground $0^+$ states. Such $E3$ transitions are also generalized seniority changing transitions ($\Delta v=2$), where the corresponding B(E3) values can be obtained as follows:
\begin{eqnarray}
B(E3)=\frac{1}{2J_i+1}| \langle {\tilde{j}}^n v l J_f || \sum_i r_i^3 Y^{3}(\theta_i,\phi_i) || {\tilde{j}}^n v\pm2 l' J_i \rangle |^2
\end{eqnarray}
The involved reduced matrix elements can similarly be obtained by using Eq.(2) between initial $J_i$ and final $J_f$ states with respective parities of $l$ and $l'$.  
\end{itemize}

\begin{figure}[!ht]
\includegraphics[width=12cm,height=11cm]{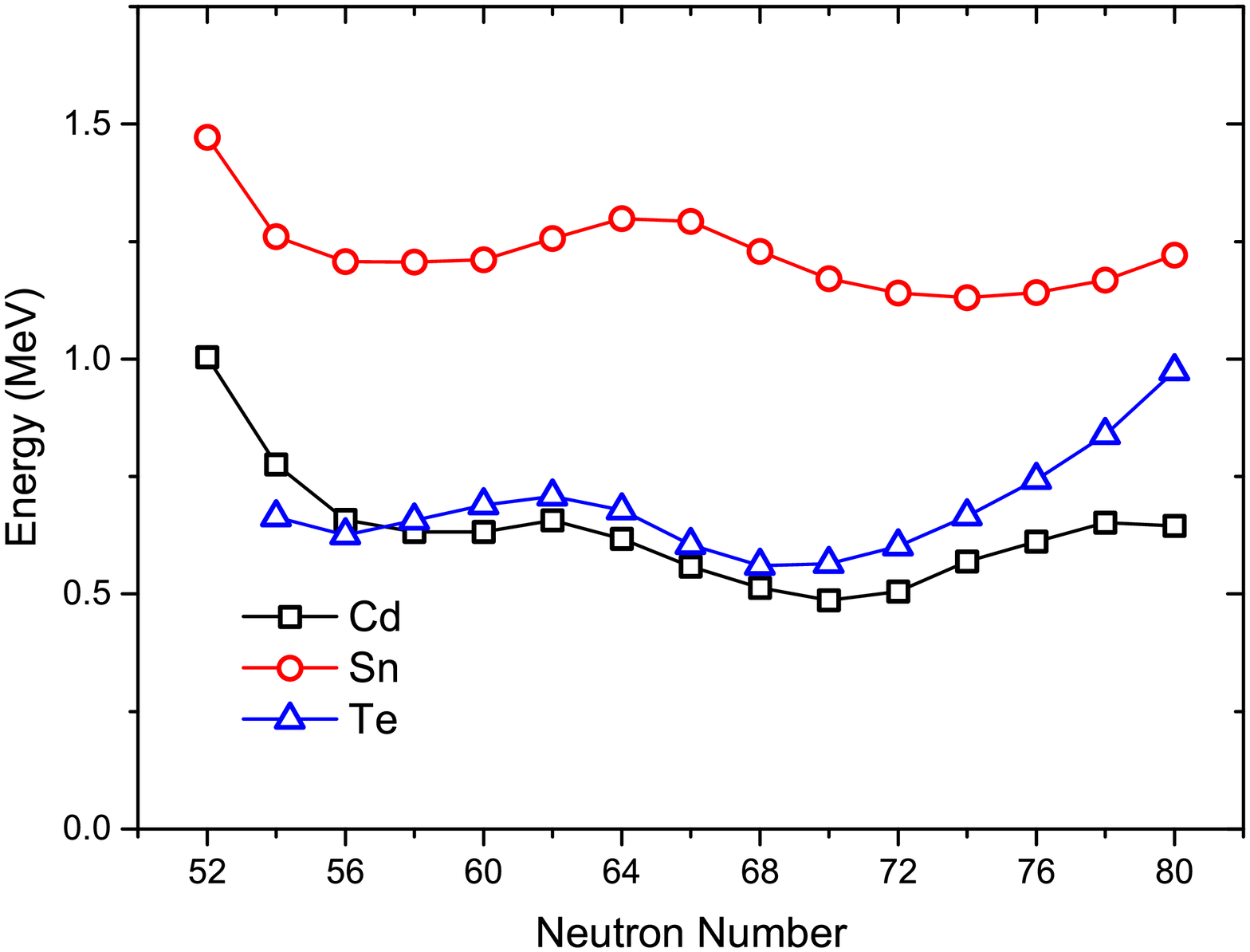}
\caption{\label{fig:energy2}(Color online) Empirical energy systematics \cite{ensdf} of first excited $2^+$ states in Cd, Sn and Te isotopic chains, respectively.}
\end{figure}

\begin{figure}[!ht]
\includegraphics[width=12cm,height=11cm]{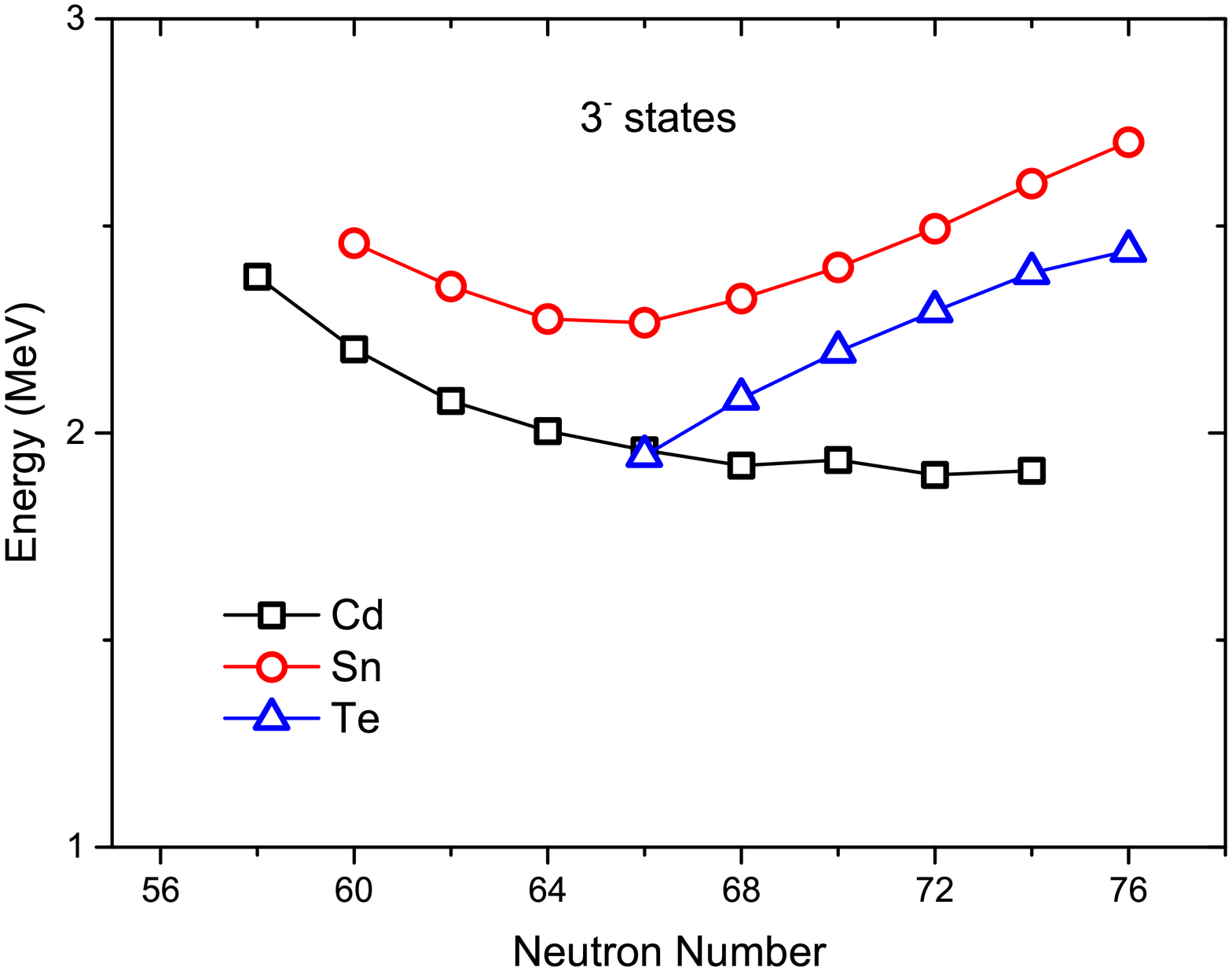}
\caption{\label{fig:energy3}(Color online) Empirical energy systematics \cite{ensdf} of first excited $3^-$ states in Cd, Sn and Te isotopic chains, respectively.}
\end{figure}

\section{Empirical support}

Figure~\ref{fig:energy2} exhibits the experimental~\cite{ensdf} energy variation of the first excited $2^+$ states with the neutron number in Cd, Sn and Te isotopes. One may note a nearly constant energy trend in all the three isotopic chains throughout $N=52-80$. An abrupt rise in the energies of Cd isotopes at $N=52, 54$ can be seen, which is very similar to case of Sn isotope at $N=52$. No experimental value is available for Te isotope at $N=52$; though, a similar case of energy rise can be seen in Te isotopes having $N \ge 76$. This may be due to approaching the respective neutron closed shell configurations at $N=50$ and $N=82$. The nearly constant energies for these $2^+$ states on going from $N=52$ to $N=80$ strongly hint towards the goodness of the generalized seniority in all the three isotopic chains. A small hump in the energy variation around middle ($N=64$) can be noticed, which hints towards a sub-shell gap in the single-particle energies of the respective neutron orbitals. The active neutron orbitals in the $N=50-82$ valence space are $g_{7/2}$, $d_{5/2}$, $d_{3/2}$, $h_{11/2}$ and $s_{1/2}$, respectively. Out of which, $g_{7/2}$ and $d_{5/2}$ lie lower in energy than the remaining $d_{3/2}$, $h_{11/2}$ and $s_{1/2}$. As soon as the neutrons start to occupy this valence space, the higher probability is to occupy $g_{7/2}$ and $d_{5/2}$; however, once these two orbitals freeze out around $N=64$, the dominance of $h_{11/2}$ can be observed. So, this small change around $N=64$ is related to the change in filling of the orbitals for the first excited $2^+$ states in Cd, Sn and Te isotopes. This feature of change at $N=64-66$ can also be followed from the shell model calculations of Sn isotopes \cite{qi2013}. Also, the same argument was given by Morales et al. \cite{morales2011} and by us \cite{maheshwari2016}, while explaining the B(E2) trends of the first $2^+$ states; which will be discussed in next section. 
These energy values in Cd and Te isotopes are consistently lower (nearly half) in comparison to the energies in Sn isotopes (with $Z=50$ closed shell), due to moving away from the closed shell. However, the generalized seniority remains constant as $v=2$ leading to a nearly particle number independent energy variation of the first excited $2^+$ states in full chain of isotopes, as shown in Fig.~\ref{fig:energy2}. 

Fig.\ref{fig:energy3} exhibits the empirical energy variation for the first excited $3^-$ states in even-even Cd, Sn and Te isotopes, wherever data are available. Again, a nearly constant variation is visible in all the three Cd, Sn and Te isotopic chains with a dip at neutron number $66$, i.e. the middle of the neutron valence space $50-82$. The energies in Cd and Te isotopes are consistently lower to the energy values in Sn isotopes. These $3^-$ states may support octupole character, if they arise from the $d-h$ orbitals of neutron valence space.  

\begin{figure}[!ht]
\includegraphics[width=12cm,height=11cm]{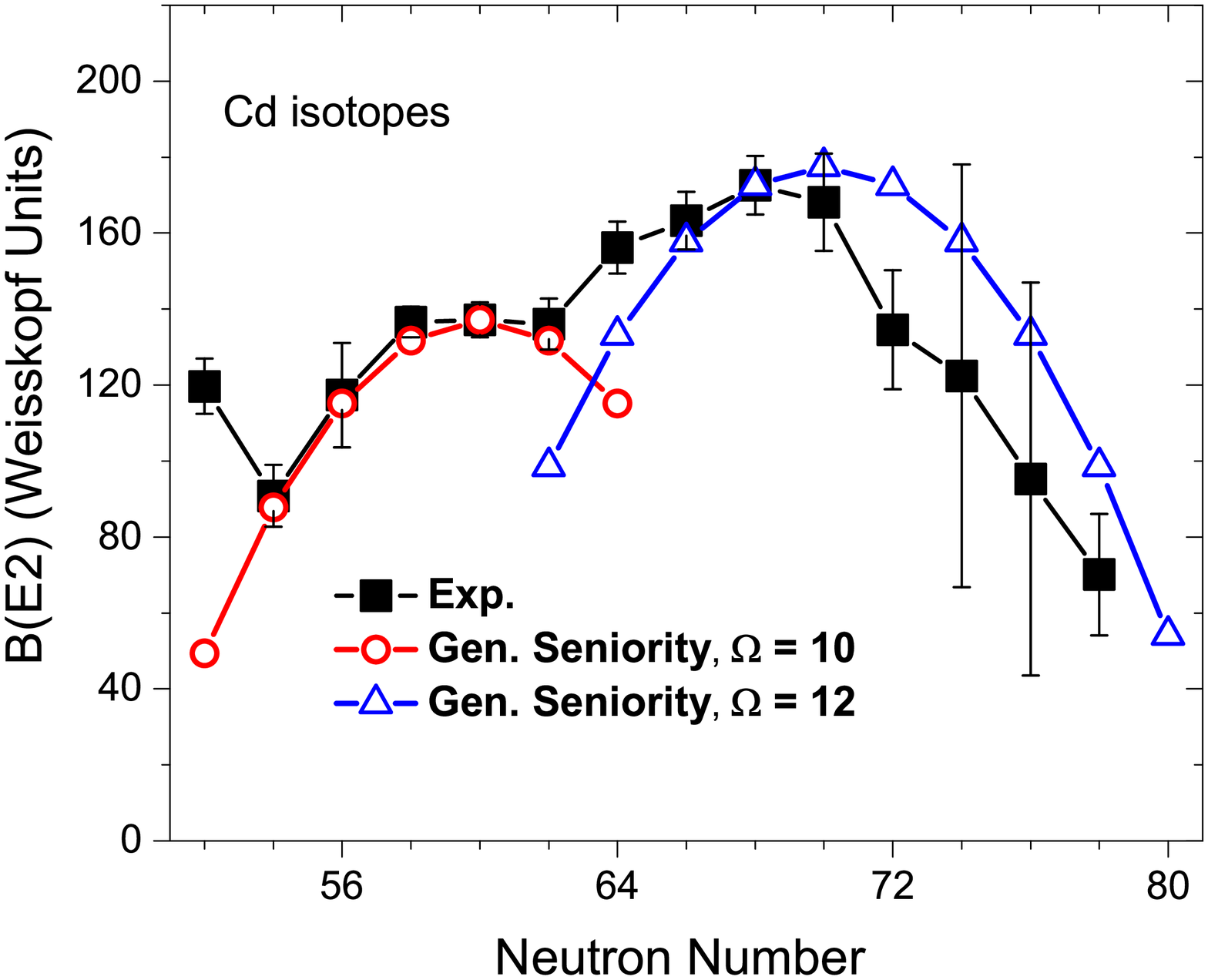}
\caption{\label{fig:be2_cd}(Color online) Comparison of the experimental~\cite{pritychenko2016} and generalized seniority calculated B(E2) trends for the first excited $2^+$ states in Cd isotopes. The asymmetry in the overall trend is explained by the filling of different orbitals before and after the mid-shell, resulting in a dip around middle. The chosen set of multi-j configuration in the generalized seniority calculations are chosen as $\Omega=10$ (before the middle) and $\Omega=12$ (after the middle), corresponding to $g_{7/2} \otimes d_{5/2} \otimes d_{3/2} \otimes s_{1/2}$, and $d_{5/2} \otimes h_{11/2} \otimes d_{3/2} \otimes s_{1/2}$, respectively.}
\end{figure}

\begin{figure}[!ht]
\includegraphics[width=12cm,height=11cm]{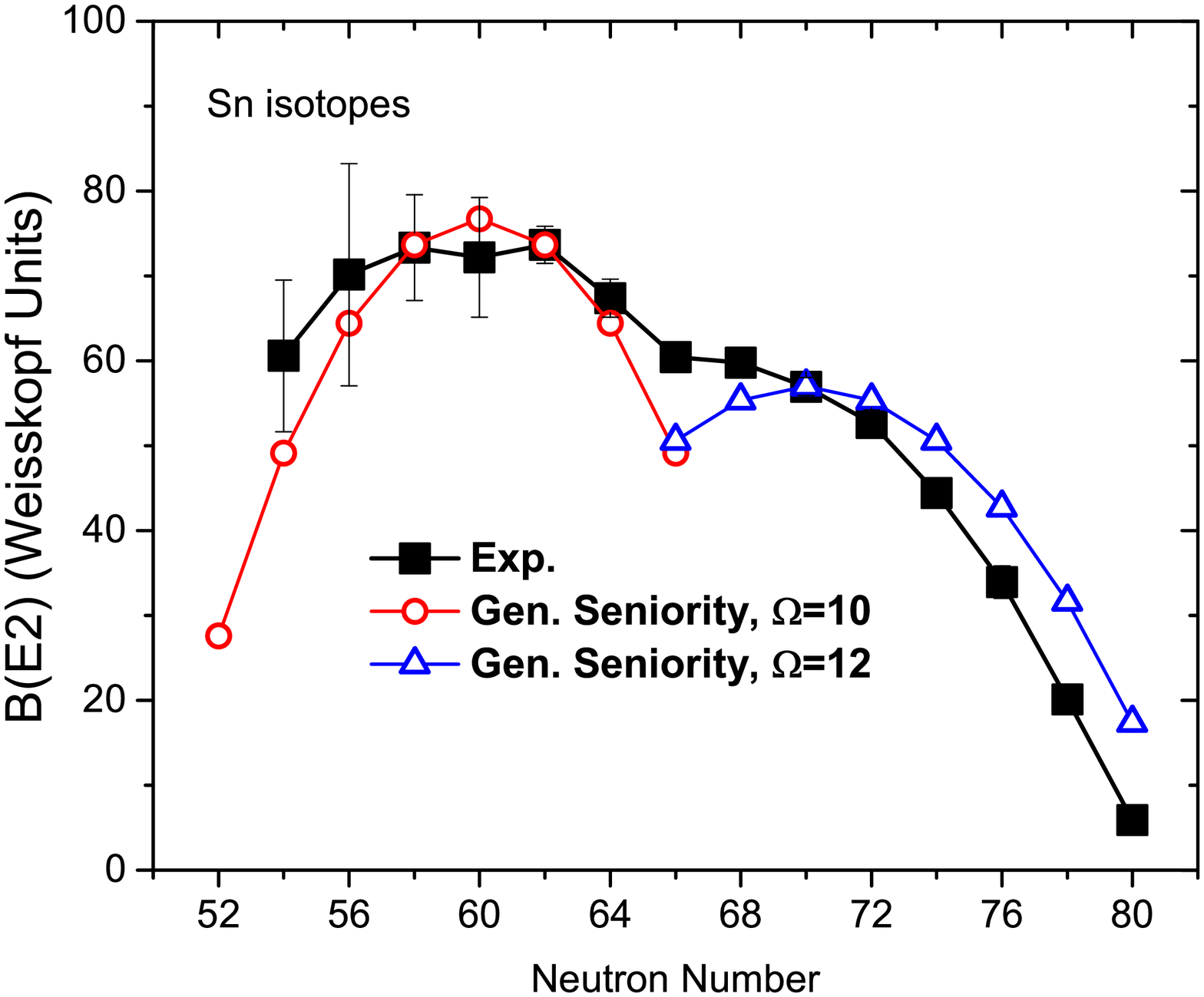}
\caption{\label{fig:be2_sn}(Color online) Same as Fig.~\ref{fig:be2_cd}, but for Sn isotopes.}
\end{figure}

\begin{figure}[!ht]
\includegraphics[width=12cm,height=11cm]{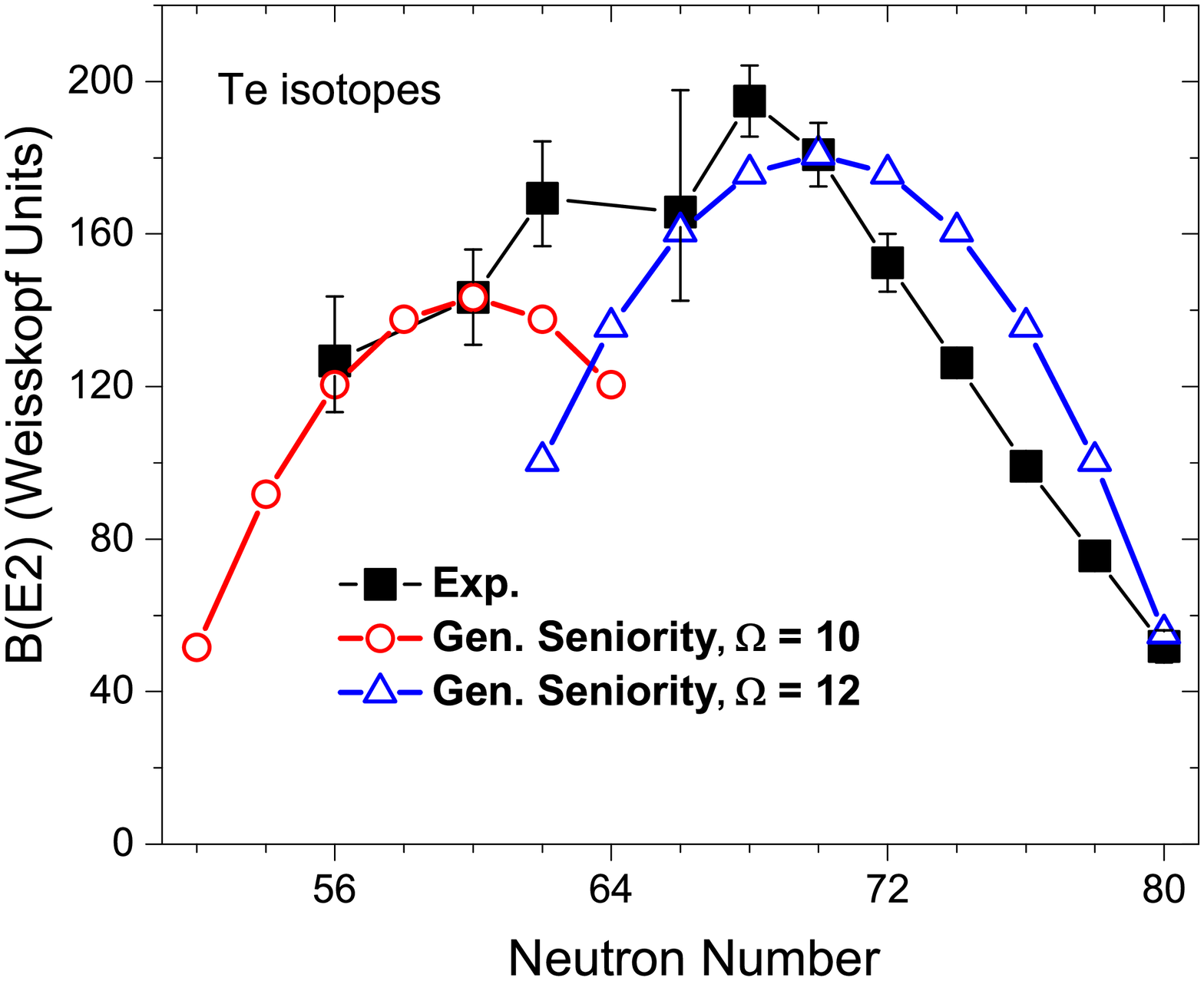}
\caption{\label{fig:be2_te}(Color online) Same as Fig.~\ref{fig:be2_cd}, but for Te isotopes.}
\end{figure}

\begin{figure}[!ht]
\includegraphics[width=12cm,height=11cm]{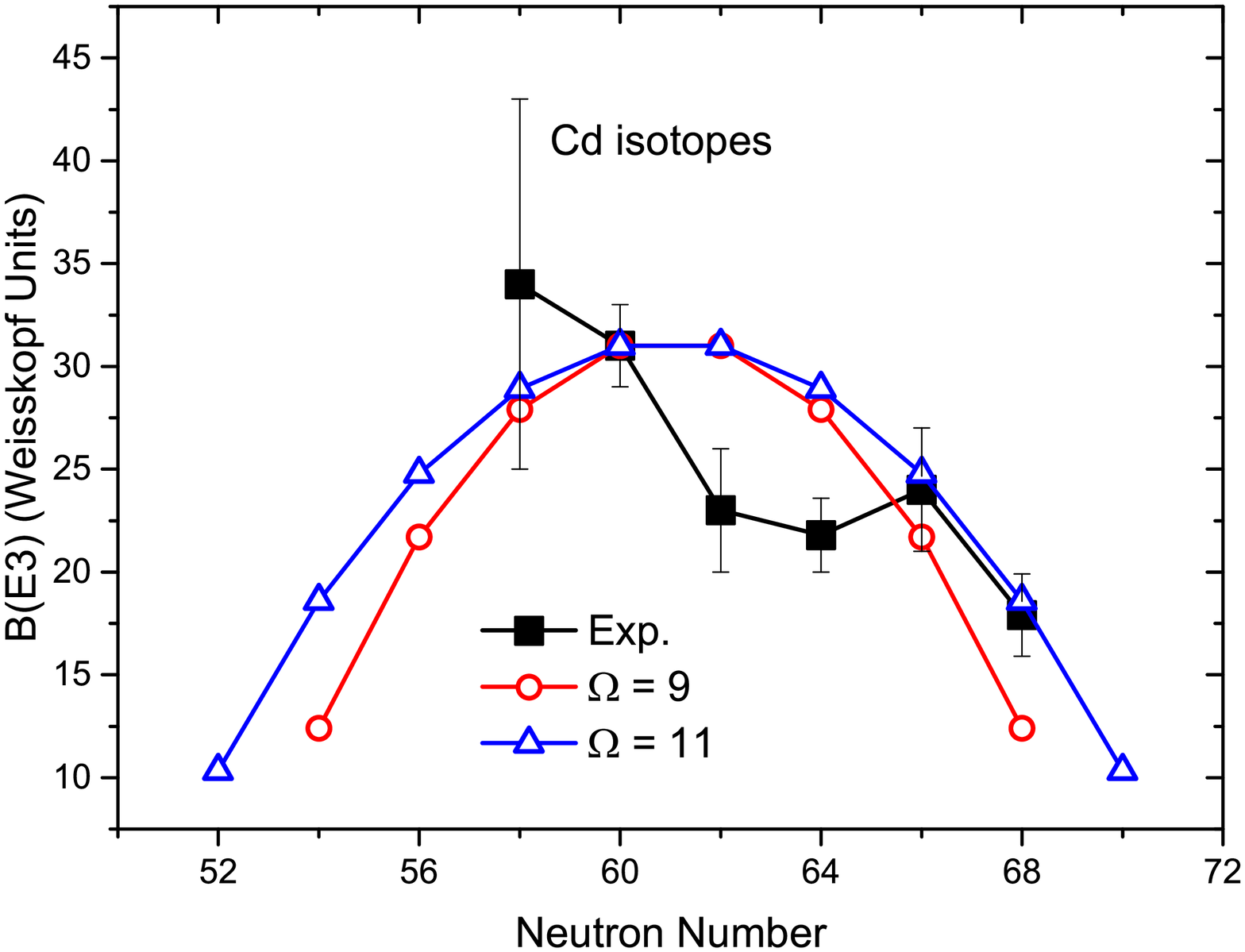}
\caption{\label{fig:be3_cd}(Color online) B(E3) variation with neutron number for the first excited $3^-$ states in Cd isotopes by using generalized seniority scheme with $\Omega=9$ and $11$ corresponding to the $d_{5/2} \otimes h_{11/2}$ and $d_{5/2} \otimes d_{3/2} \otimes h_{11/2}$ configurations, respectively.}
\end{figure}

\begin{figure}[!ht]
\includegraphics[width=12cm,height=11cm]{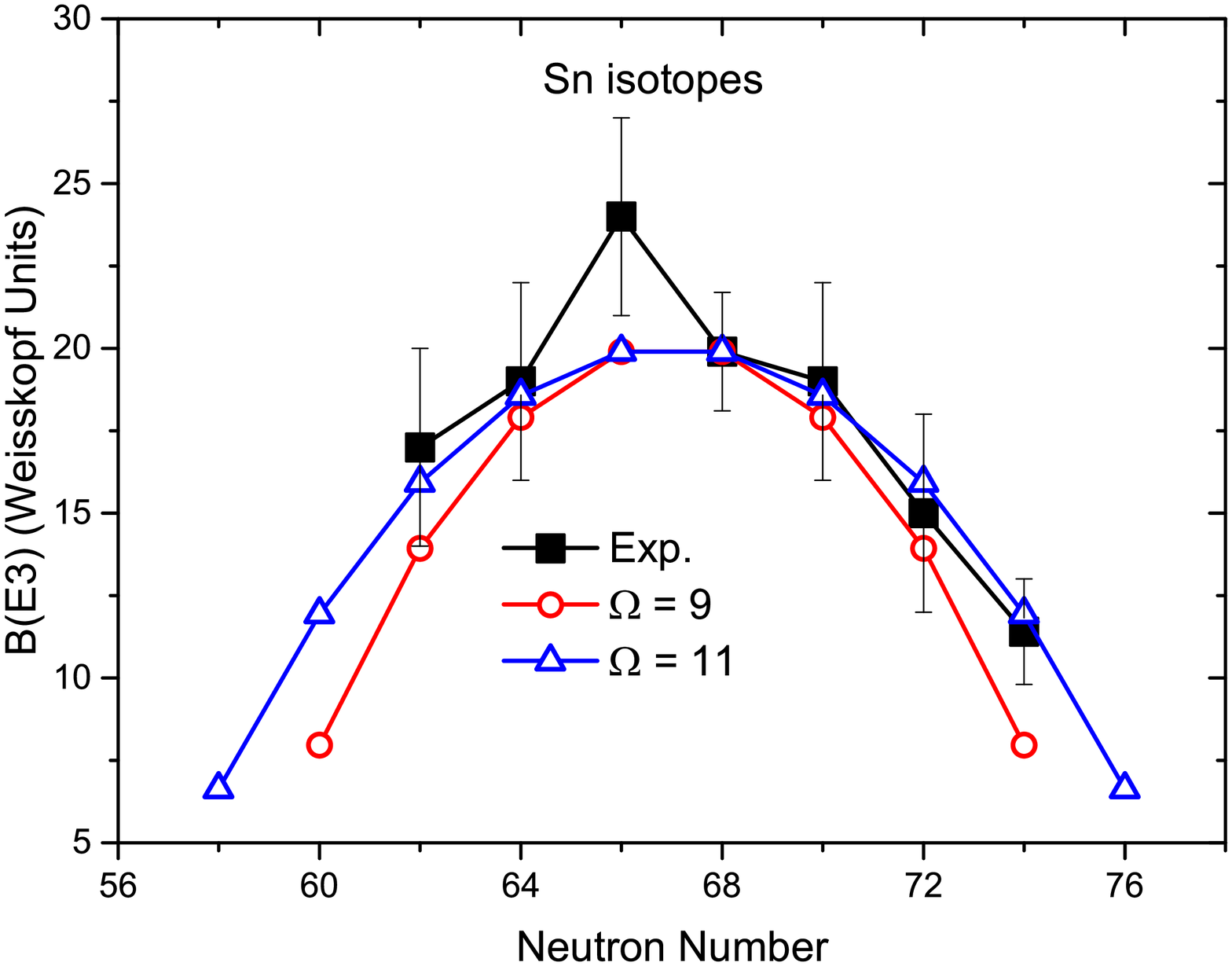}
\caption{\label{fig:be3_sn}(Color online) Same as Fig.~\ref{fig:be3_cd}, but for the Sn isotopes.}
\end{figure}

\begin{figure}[!ht]
\includegraphics[width=12cm,height=11cm]{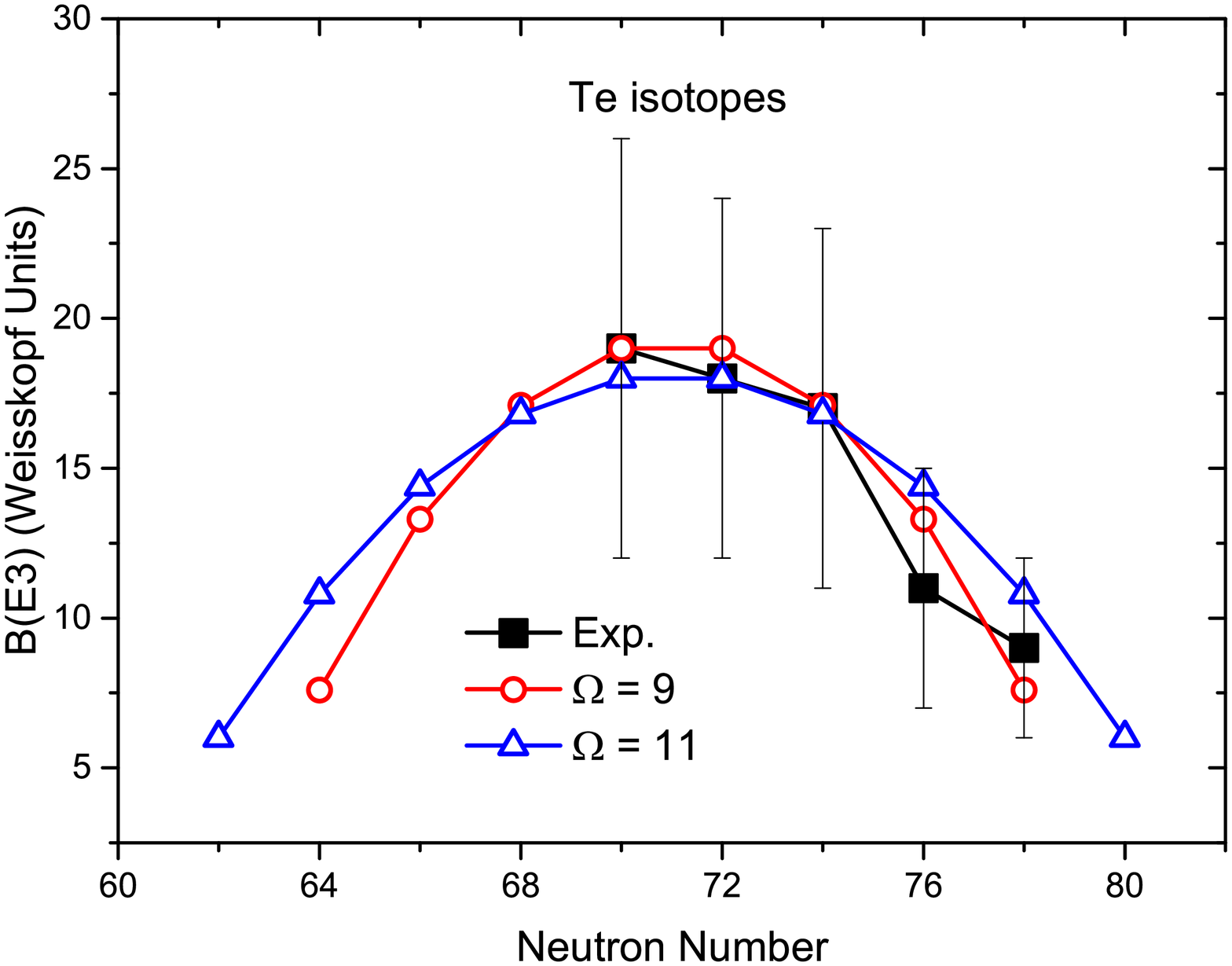}
\caption{\label{fig:be3_te}(Color online) Same as Fig.~\ref{fig:be3_cd}, but for Te isotopes.}
\end{figure}

\section{\label{sec:level4}Results and Discussion}

In this section, we discuss the generalized seniority calculated results for the first excited $2^+$ states and ${3}^-$ states in even-A Cd, Sn and Te isotopes. The pair degeneracies $\Omega=9, 10, 11$ and $12$ correspond to the configurations \{$ d_{5/2} \otimes h_{11/2}$\}, \{$g_{7/2} \otimes d_{5/2} \otimes d_{3/2} \otimes s_{1/2}$\}, \{$ d_{5/2} \otimes d_{3/2} \otimes h_{11/2}$\}, and \{$d_{5/2} \otimes h_{11/2} \otimes d_{3/2} \otimes s_{1/2}$\}, respectively, in the following discussion.

\subsection{Asymmetric double-hump behavior of B(E2) values}

First we review our results on two asymmetric B(E2) parabolas for the first excited $2^+$ states in Cd, Sn and Te isotopes \cite{maheshwari20161,maheshwari20192}. Recent B(E2) evaluation~\cite{pritychenko2016} has been used to obtain the experimental systematic trends of the B(E2) values for Cd and Te isotopes. The evaluated B(E2) values for Sn isotopes is used from ~\cite{maheshwari20161}. We note a nearly identical pattern of two asymmetric B(E2) parabolas for all the three Cd, Sn and Te isotopic chains as shown in Figs. \ref{fig:be2_cd}, ~\ref{fig:be2_sn} and ~\ref{fig:be2_te}, respectively. A dip around the middle in the experimental B(E2) values is also visible for all the three isotopic chains. 

The generalized seniority calculations use the multi-j configuration corresponding to $\Omega=10$ (before the middle) and $\Omega=12$ (after the middle), respectively. The active set of orbitals is mainly dominated by $g_{7/2}$ and $d_{5/2}$ orbitals before the middle, while the $h_{11/2}$ orbital dominates after the middle. The calculated trends depend on the square of the coefficients in Eq.(2), since the $0^+$ to $2^+$ transitions are generalized seniority changing $\Delta v=2$ transitions. To take care of other structural effects, we fit one of the experimental data and restrict the values of radial integrals and involved $3j-$ and $6j-$ coeffcients as a constant, which should be the case for an interaction conserving the generalized seniority. 

It is interesting to note that the generalized seniority calculated values explain the overall trends of the experimental data in all the three Cd, Sn and Te isotopic chains, see Figs.~\ref{fig:be2_cd}, ~\ref{fig:be2_sn} and ~\ref{fig:be2_te}, respectively. Asymmetry in the inverted parabola before and after the middle has again been attributed to the difference in filling the two sets of orbitals. The dominance of $g_{7/2}$ orbital gets shifted to the $h_{11/2}$ orbital near the middle of the shell resulting in a dip. However, the generalized seniority remains constant at $v=2$ leading to the particle number independent energy variation for the $2^+$ states throughout the full chain. The generalized seniority, hence, governs the electromagnetic properties not only in Sn isotopes but also in the Cd and Te isotopes, which are not semi-magic nuclei. One may note that the calculations only consider the active orbitals of $N=50-82$ valence space. No signs of shell quenching have, therefore, been witnessed for these first excited $2^+$ states in all the three Cd, Sn and Te isotopes. However, we note that the influence of two proton holes/ particles cannot be ignored; the overall trend is largely obtained by changing the neutron number in total E2 transition matrix elements. 

\subsection{Inverted parabolic behaviour of B(E3) values}

Figs.~\ref{fig:be3_cd}, ~\ref{fig:be3_sn} and ~\ref{fig:be3_te} exhibit the experimental and generalized seniority calculated B(E3) trends for Cd, Sn and Te isotopic chains. The experimental data are taken from the compilation of Kibedi and Spear ~\cite{kibedi2002}. The calculations are done by using the pair degeneracy of $\Omega=9$ and $11$ corresponding to the $d_{5/2} \otimes h_{11/2}$ and $d_{5/2} \otimes d_{3/2} \otimes h_{11/2}$ configurations, respectively. An inverted parabolic B(E3) trend is visible in all the three Cd, Sn and Te isotopic chains. The calculated results explain the gross experimental trend in the best possible way, wherever available. However, one can notice that the Fermi surface for the given configurations seems to be different for each chain. Therefore, the peaks belong to different neutron number in each chain, such as $N=58-60$ for Cd isotopic chain, $N=66$ for Sn isotopic chain and $N=70$ for Te isotopic chain.  The deviation of many points from the generalized seniority are more pronounced in Cd isotopes. More precise measurements are needed to improve/test the theoretical arguments.  

\section{\label{sec:level5}Summary and Conclusion}

We have revisited the B(E2) and B(E3) trends for the first excited $2^+$ and $3^-$ states in Cd, Sn and Te isotopes and examined them on the basis of the multi-j generalized seniority scheme. Two asymmetric B(E2) parabolas for the first $2^+$ states have been noticed in Cd (two-proton holes) and Te (two-proton particles) isotopes, and explained in terms of generalized seniority identical to the case of Sn isotopes. The consistency of configuration on going from two-holes to two-particles is remarkable, which results in a nearly particle number independent energy variation for the first $2^+$ states in all the three isotopic chains. 

The generalized seniority scheme further explains the inverted parabolic B(E3) trend for the first excited $3^-$ states, not only in Sn isotopes but also in Cd and Te isotopes, for the first time. New and more precise measurements are needed to confirm these results, particularly in Cd and Te isotopes. No shell quenching is suggested for these low lying states. The role of the symmetries related to pairing shows up in the goodness of generalized seniority quantum number, which offers a unified view of the $2^+$ and $3^-$ states in Cd, Sn and Te isotopes. 

\section{Acknowledgement}

The author would like to thank Prof. A. K. Jain for innumerable thought-provoking discussions. The work was done during the author's post-doctorate at University of Malaya. 



\end{document}